\documentclass[structabstract]{aa}
\usepackage{graphicx}
\usepackage{amsmath}
\usepackage[varg]{txfonts}
\usepackage{xcolor}
\usepackage{hyperref}

\newcommand{\polsnr}{\ensuremath{\mathrm{polSNR}}}
\usepackage{orcidlink}

\begin{document}

\title{Parallel and perpendicular: a two-population test with POSSUM radio galaxies confirms hint of CMB birefringence}

\author{M. Br\"uggen\inst{1}\orcidlink{0000-0002-3369-7735}
   \and J. Dreismann\inst{1}\orcidlink{0009-0003-4099-8565}
   \and P. Trivedi\inst{1}\orcidlink{0000-0001-5748-696X}
   \and S. P. O'Sullivan\inst{2}\orcidlink{0000-0002-3968-3051}
   \and T. Akahori\inst{3,4}\orcidlink{0000-0001-9399-5331}
   \and F. Naokawa\inst{2,5,6}\orcidlink{0000-0003-3542-6637}
   \and C.L.~Van Eck\inst{7}\orcidlink{0000-0002-7641-9946}
   \and M. Regis\inst{8,9}\orcidlink{0000-0003-0399-0284}
   }

\institute{Hamburger Sternwarte, Universit\"at Hamburg, Gojenbergsweg 112, 21029 Hamburg, Germany\\
  \email{marcus.brueggen@uni-hamburg.de}
  \and Departamento de F\'isica de la Tierra y Astrof\'isica \& IPARCOS-UCM, Universidad Complutense de Madrid, 28040 Madrid, Spain
  \and Mizusawa VLBI Observatory, National Astronomical Observatory of Japan, 2-21-1 Mitaka, Tokyo 181-8588, Japan
  \and SOKENDAI, Shonan Village, Hayama-machi, Miura-gun, Kanagawa 240-0193 Japan
  \and Department of Physics, Tohoku University, Sendai, Miyagi 980-8578, Japan
  \and Research Center for the Early Universe, The University of Tokyo, Bunkyo-ku, Tokyo 113-0033, Japan
  \and Research School of Astronomy \& Astrophysics, The Australian National University, Canberra, ACT 2611, Australia
  \and Dipartimento di Fisica, Universit\`{a} di Torino, via P. Giuria 1, I--10125 Torino, Italy
  \and INFN, Sezione di Torino, via P. Giuria 1, I--10125 Torino, Italy
  }

\date{Received ...; accepted ...}

\abstract
{Several analyses of the polarisation of the cosmic microwave background (CMB) find an isotropic rotation of the plane of linear polarisation, cosmic birefringence, of $\beta\sim0.2^\circ$--$0.3^\circ$, which hints at parity-violating physics. An independent test at low redshift and with different systematics is important to verify these results.}
{Extended radio galaxies provide such a test because their Faraday de-rotated polarisation angle $\chi_0$ is statistically aligned with their orientation on the sky, measured through the position angle $\psi$. Previous tests fitted only the population with $\chi_0-\psi\approx90^\circ$. We add the population with $\chi_0-\psi\approx0^\circ$, which would be rotated by the same angle.}
{We fit the distribution of $\chi_0-\psi$ with two axial von-Mises distributions plus a uniform background, using an unbinned likelihood that includes the angle error of each source. We apply this to 33,298 components of 16,649 double radio sources from the Polarisation Sky Survey of the Universe's Magnetism (POSSUM), in which both components have a polarised signal-to-noise ratio $>10$ and are separated by more than half the beam.}
{The $0^\circ$ population is as abundant as the $90^\circ$ population. In the CMB sign convention we measure $\beta=0.48^\circ\pm0.35^\circ$, in agreement with the CMB in sign and amplitude. Including the $0^\circ$ population halves the variance of $\beta$ compared with a fit of the $90^\circ$ population alone. Components closer than half a beam contain no information on the source axis. Excluding them leaves $\beta$ the same and makes the two populations consistent in every morphological class. Combining with the joint ACT and \textit{Planck} likelihood under the hypothesis of a late-time rotation yields $\beta=0.282^\circ\pm0.056^\circ$, $5.0\sigma$ from zero.}
{Using both polarisation populations doubles the information in radio-galaxy birefringence tests. At the typical redshift of radio galaxies, $z\sim0.5$, the test probes the rotation accumulated over the last $\sim5$\,Gyr. A rotation of the CMB size at these redshifts would point to a field that evolves at late times, such as an axion-like dark-energy field, whereas a rotation that occurred before reionisation, for example by an axion-like dark-matter field, predicts $\beta\approx0$. With control of the absolute angle calibration, the completed POSSUM survey can test the CMB result independently.}

\titlerunning{Two-population test of cosmic birefringence}
\authorrunning{M. Br\"uggen et al.}
\maketitle

\section{Introduction}

Cosmic birefringence denotes a rotation of the plane of linear polarisation of light due to different phase velocities of the two circular-polarisation components. It can arise, for instance, if photons couple to a background four-vector $p_\mu$ through a Chern--Simons term $\tfrac12 p_\mu A_\nu\tilde F^{\mu\nu}$ \citep{Carroll1990}. A constant timelike $p_\mu$ violates Lorentz and CPT invariance and produces a rotation that grows with distance \citep{KosteleckyMewes2008}. If $p_\mu=g\,\partial_\mu\phi$ is the gradient of an axion-like field, the rotation tracks the change of the field between emission and observation \citep{Harari1992,Carroll1998}. Such a field could be part of dark matter or dark energy and induces a cosmic birefringence that is, unlike Faraday rotation, independent of frequency \citep{Choi2021,Fujita2021,Komatsu2022}.

Analyses of CMB polarisation that calibrate the instrumental angles with Galactic foregrounds or optics models report $\beta=0.342^{+0.094}_{-0.091}$\,deg from WMAP and \textit{Planck} \citep{MinamiKomatsu2020, DiegoPalazuelos2022, Eskilt2022, EskiltKomatsu2022} and $0.215^\circ\pm0.074^\circ$ from ACT DR6 \citep{DiegoPalazuelos2025, Louis2025}. A joint analysis of ACT and \textit{Planck} finds $0.277^\circ\pm0.057^\circ$ \citep{Eskilt2026}, while noting that unresolved systematics remain. An independent test with different instruments and at low redshift is therefore essential to distinguish among different interpretations.

In powerful double radio sources, the lobe magnetic field is on average parallel to the source axis. The polarisation is then perpendicular to it, so $\Delta\equiv\chi_0-\psi$ peaks at $90^\circ$ \citep{Gardner1969,HavesConway1975,Clarke1980}. A rotation by $\beta$ shifts this peak as explored in the 90s \citep{Carroll1990,CarrollField1991}. 
\citet{Trivedi2026} used this with edge-brightened POSSUM sources to find $\beta=-0.19^\circ\pm0.72^\circ$ in the IAU convention, that is, $+0.19^\circ$ in the CMB convention (Sect.~\ref{sec:systematics}). \citet{Battye2026} reached similar precision with compact sources. These analyses fit one peak plus a uniform background that contains 70--80\% of the sources. However, a second population, polarised parallel to the axis, is well known from hot spots and jets where the plasma flows change direction \citep{SaikiaSalter1988,ListerHoman2005}, and it is rotated in exactly the same way. Here we show that this population is as abundant as the standard crosses and use both in a joint likelihood. To our knowledge this has not been done before.

\section{Data and sample}
\label{sec:data}
The Polarisation Sky Survey of the Universe's Magnetism (POSSUM; \citealt{Gaensler2025}) images the sky at 800--1088\,MHz with the Australian Square Kilometre Array Pathfinder (ASKAP; \citealt{Hotan2021}) at $\sim20''$ resolution and $\sim18\,\mu$Jy\,beam$^{-1}$ in Stokes $Q$ and $U$, commensally with the Evolutionary Map of the Universe (EMU; \citealt{Norris2021}).
 \textsc{Selavy} \citep{Whiting2012} decomposes sources into islands and Gaussian components. For each component we use the polarised intensity $P$, \polsnr, and the angle $\chi_0$. 

Faraday rotation in magnetised plasma along the line of sight rotates the polarisation angle as $\chi(\lambda^2)=\chi_0+\mathrm{RM}\,\lambda^2$, with the rotation measure $\mathrm{RM}=0.812\int n_\mathrm{e}B_\parallel\,\mathrm{d}l$\,rad\,m$^{-2}$ ($n_\mathrm{e}$ in cm$^{-3}$, $B_\parallel$ in $\mu$G, $l$ in pc). The catalogue gives the RM from RM synthesis and the angle $\chi_0=\chi-\mathrm{RM}\,\lambda_\mathrm{ref}^2$, Faraday de-rotated to $\lambda=0$, with an uncertainty $\sigma_i$. Before use, we correct the catalogued angles for the reference-frame error described in Sect.~\ref{sec:systematics}. We use the Data Release 1 precursor catalogue of May 2026 ($\sim40\%$ of the survey, van Eck et al. (in prep.) and Anderson et al. (in prep.)), de-duplicated as in \citet{Trivedi2026}. Excluding single-component islands and $|b|<10^\circ$ leaves 791,725 components in two-component islands. 

With a catalogue of this size, visual inspection cannot practically identify all edge-brightened radio galaxies.
Hence, we decided to use a machine-learning morphological classification. For each source from our catalogue, we first extract an EMU image cutout centred on its mean catalogue position, then scale, sigma-clip, and normalise it so that all inputs to the network are uniform. The cutouts are then classified with a self-supervised deep-clustering pipeline that was developed by \citet{Perez2025}, originally for the Low Frequency ARray (LOFAR). It groups the unlabelled sources into morphological classes and needs no large, labelled training set.


In the self-supervised classification of sources from the second data release of the LOFAR Two-metre Sky Survey (LoTSS-DR2) by \citet{Perez2025}, classes 7, 8 and 9 comprise the symmetric double, edge-bright and diffuse-bent sources.

Symmetric doubles fearure two jets and form by far the most populous class.
Edge-brightened sources show bright hotspots, in some cases with a core and lobes. Symmetric doubles and edge-bright sources together constitute the FRII sources, which the model separates well from FRI and bent sources. Here $\psi$ is the position angle of the line joining the two components. For bent sources, $\psi$ traces the mean axis rather than the local jet direction. Bending adds scatter to $\Delta$, which lowers $\kappa$, but it does not shift the peaks as long as sources bend equally often in both directions. The angles $\chi_0$ and $\psi$ follow the IAU convention, increasing from north through east \citep{HamakerBregman1996}.

The median $\sigma_i$, which includes the extrapolation of the rotation measure (RM) to $\lambda=0$, is $44^\circ$, $22^\circ$ and $7^\circ$ for $3<\polsnr<4$, $6<\polsnr<10$ and $\polsnr>10$. No peak in the distribution of $\Delta$ is detectable below $\polsnr=10$, and these components would receive negligible weight, so we require $\polsnr>10$. A significant $90^\circ$ peak is present in classes 7, 8 (the edge-brightened class of \citealt{Trivedi2026}) and 9, but not in the others. In these classes, 61,938 components in 40,199 islands pass the cut. 

The structural angle $\psi$ traces the source axis only if the two components are resolved from each other. Thus, for our baseline sample we require that both components lie above the cut and are separated by more than $10''$, which is half the restoring beam. This leaves 33,298 components in 16,649 sources (6,254, 3,484 and 23,560 in classes 7, 8 and 9, respectively), with a median $P=0.67$\,mJy. In Sect.~\ref{sec:results} we show that the cut removes components that carry no information on the axis, thus leaving $\beta$ unchanged.

\section{Method}
\label{sec:method}

If the projected field is parallel to the axis, as in radio galaxy lobes, $\Delta \approx 90^\circ$. If it is perpendicular, as in compressed hot spots or toroidal jets, $\Delta\approx0^\circ$. Islands whose components are not the two lobes can contribute to either population. Each component enters the likelihood with its own population probabilities, so a source whose two components belong to different populations is included naturally.

A rotation that is the same for all sources shifts both peaks by the same $\beta$. If the rotation grows with distance, as for a constant Lorentz-violating $p_\mu$ \citep{Carroll1990} or dynamical axion-like particles \citep{Namikawa2025, 2026PhRvL.136d1004N} , each peak measures $\beta$ averaged over the redshift distribution of its own population, and the two agree only if these distributions are the same. The peak difference $\beta_{90}-\beta_0$ (Sect.~\ref{sec:results}) tests this. 

For a rotation that matches the CMB value and scales with look-back time \citep{2026PhRvL.136d1004N}, the expected difference is below $0.1^\circ$, far below our precision. 

With the axial von Mises density
\begin{equation}
\mathcal{V}(\Delta;\mu,\kappa)=\frac{\exp[\kappa\cos2(\Delta-\mu)]}{\pi I_0(\kappa)}
\end{equation}
\citep{vonMises1918,MardiaJupp2000}, where $\mu$ is the mean direction (the peak position of $\Delta$, defined modulo $180^\circ$), $\kappa\ge0$ is the concentration ($\kappa=0$ gives a uniform distribution. For large $\kappa$ the density approaches a Gaussian of width $\sigma_\Delta\simeq1/(2\sqrt\kappa)$), and $I_n$ is the modified Bessel function of the first kind of order $n$, the probability density of component $i$ is

\begin{equation}
p(\Delta_i)=f_i\,\mathcal{V}(\Delta_i;\mu,\kappa_i)+f_0\,\mathcal{V}(\Delta_i;\mu+90^\circ,\kappa_{0,i})+\frac{1-f_i-f_0}{\pi},
\label{eq:mixture}
\end{equation}
with $\mu=90^\circ+\beta$. The fractions are given by:

\begin{equation}
f_i=\frac{e^{a+bx_i}}{1+e^{a+bx_i}+e^{c}},\quad f_0=\frac{e^{c}}{1+e^{a+bx_i}+e^{c}},\quad x_i=\log_{10}\frac{P_i}{0.5\,\mathrm{mJy}}.
\end{equation}
\citet{Trivedi2026} instead weighted each angle below $P=0.5$\,mJy by $(P/0.5\,\mathrm{mJy})^2$. Here every component enters with equal weight, and the data determine how the probability of belonging to the $90^\circ$ population increases with $P$. Faint components then contribute mainly to the uniform background.

Convolving a von-Mises distribution of the doubled angle with a wrapped normal distrbution of width $2\sigma_i$ multiplies its mean resultant length $A(\kappa)=I_1(\kappa)/I_0(\kappa)$ by $e^{-2\sigma_i^2}$ \citep{MardiaJupp2000}. Each component thus has an effective concentration given by $A(\kappa_i)=A(\kappa)\,e^{-2\sigma_i^2}$, and likewise for $\kappa_{0,i}$. We invert $A$ following \citet{BestFisher1981} and maximise $\ln\mathcal{L}=\sum_i\ln p(\Delta_i)$ over the variables $(\mu,\kappa,\kappa_0,a,b,c)$ without binning. The uncertainties are computed from the likelihood curvature, the profile likelihood and a bootstrap over islands. Because the two components of a source are not independent, we quote the bootstrap uncertainty.

On 100 mock catalogues drawn from the best-fitting model with the observed $\sigma_i$ and $P_i$ (class-8 sample, $\polsnr>6$), the estimator is unbiased to within $0.15^\circ$. Its scatter is $0.57^\circ$, compared to $0.84^\circ$ for the two-step fit of \citet{Trivedi2026}. Injecting rotations of up to $5^\circ$ into the combined sample recovers them to within $0.001^\circ$ (Sect.~\ref{sec:systematics}).

\section{Results}
\label{sec:results}

\begin{figure*}
\centering
\includegraphics[width=0.6\textwidth]{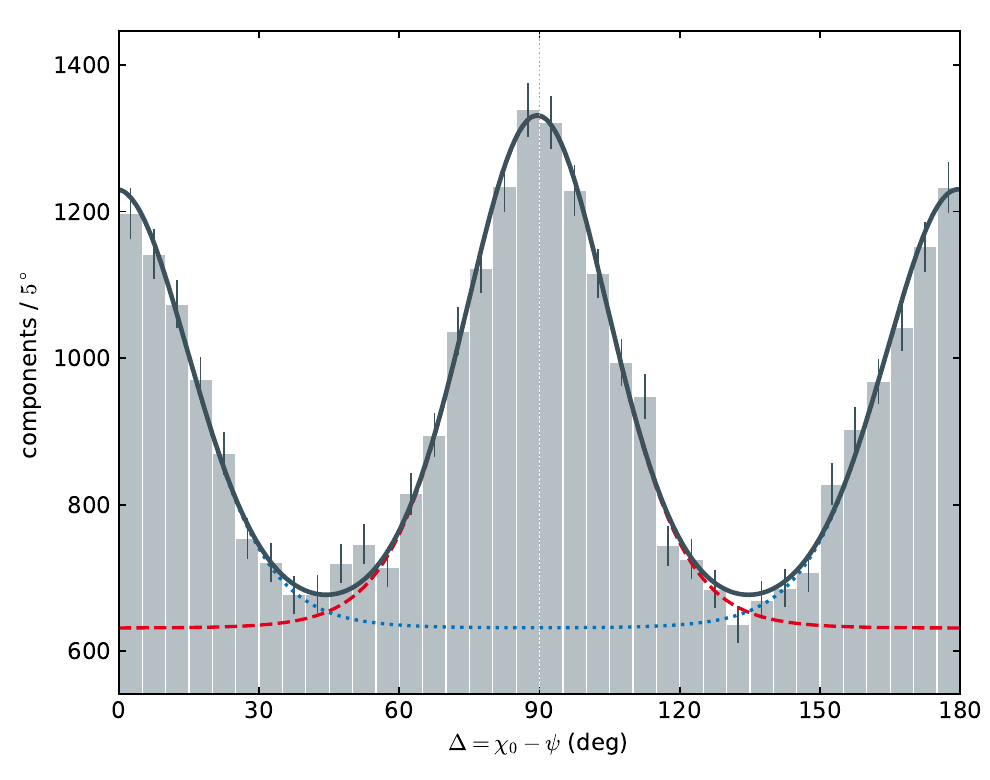}
\includegraphics[width=0.35\textwidth]{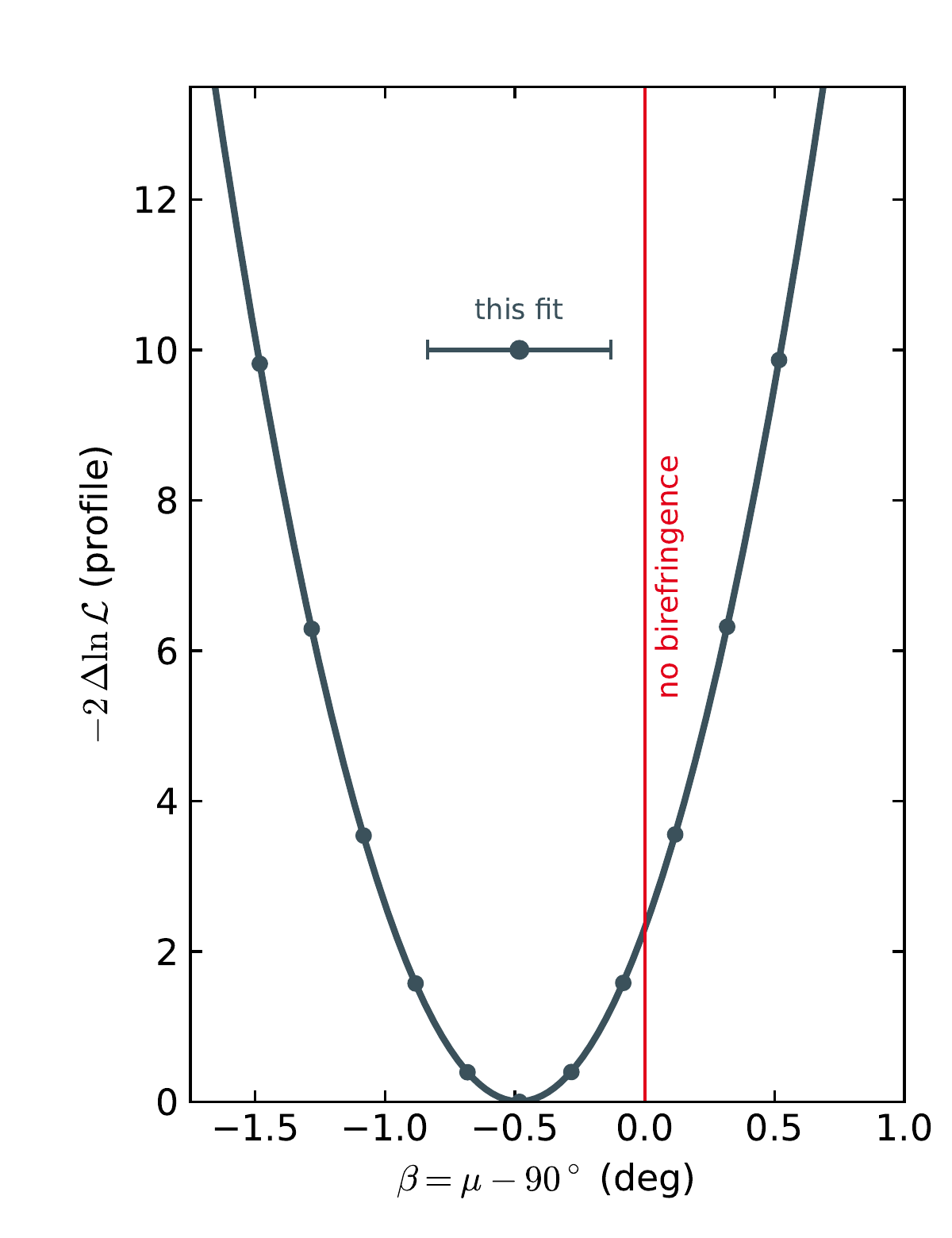}
\caption{\emph{Left:} $\Delta=\chi_0-\psi$ for the $33,298$ components of the baseline sample ($5^\circ$ bins). Curves show the unbinned model of Eq.~(\ref{eq:mixture}) summed over all components (solid), and its standard-angle (dashed) and perpendicular (dotted) parts, each added to the uniform background.  \emph{Right:} Profile likelihood of the birefringence angle $\beta$ (IAU convention).}
\label{fig:hist}
\end{figure*}

Both peaks are clearly detected, and, interestingly, with comparable amplitudes (Fig.~\ref{fig:hist}). The model describes the binned data well with a $\chi^2=28.0$ for 36 bins. We find $\kappa=3.91\pm0.23$ and $\kappa_0=3.92\pm0.30$, an intrinsic dispersion of $\sim 16^\circ$ for both populations. The standard-angle fraction increases from 0.13 at a polarised intensity of $P=0.5$\,mJy to 0.37 at $P=10$\,mJy.
This corresponds to about 5,600 standard-angle and 5,000 perpendicular components. The profile likelihood is parabolic to 1\% within $\pm1.2^\circ$. The curvature and the profile give $\sigma_\beta=0.32^\circ$. The island bootstrap (483 resamples) gives $0.35^\circ$, larger because the angles of the two components of a source are correlated. We adopt
\begin{equation}
\beta=-0.48^\circ\pm0.35^\circ\quad\text{(IAU)},
\end{equation}
which is 2 times more precise than \citet{Trivedi2026}.

\begin{table}
\caption{Fits for individual and combined classes of the baseline sample ($\polsnr>10$, separation $>10''$- Angles in degrees and using IAU convention). $\beta$ is from the joint fit. $\beta_{90}$ and $\beta_0$ from fits with independent peak centres. Uncertainties are computed from the likelihood curvature.}
\label{tab:classes}
\centering
\small
\resizebox{\columnwidth}{!}{%
\begin{tabular}{lrccc}
\hline\hline
Class & $N$ & $\beta$ & $\beta_{90}$ & $\beta_0$  \\
\hline
7     &  6,254 & $-1.00\pm0.77$ & $-1.59\pm0.86$ & $+1.28\pm1.58$  \\
8     &  3,484 & $-0.49\pm0.70$ & $-0.50\pm0.70$ & $-1.3\pm5.6$  \\
9     & 23,560 & $-0.48\pm0.39$ & $-0.17\pm0.58$ & $-0.70\pm0.50$  \\
\hline
7+8+9 & 33,298 & $-0.48\pm0.32$ & $-0.51\pm0.41$ & $-0.44\pm0.50$ \\
\hline
\end{tabular}}
\end{table}

Giving the $0^\circ$ peak its own centre yields $\beta_{90}=-0.51^\circ\pm0.41^\circ$ and $\beta_0=-0.44^\circ \pm0.50^\circ$. The difference, $-0.08^\circ\pm0.63^\circ$ including the covariance, is consistent with zero. Each population carries about half of the information. Within the individual classes the peaks differ by $1.7\sigma$, $0.1\sigma$ and $0.7\sigma$ (Table~\ref{tab:classes}), together $2\Delta\ln\mathcal{L}=3.2$ for three degrees of freedom. 

\begin{table}
\caption{Dependence on component pairing and separation ($\polsnr>10$. Angles in degrees and using IAU convention).}
\label{tab:sep}
\centering
\small
\resizebox{\columnwidth}{!}{%
\begin{tabular}{lrccc}
\hline\hline
Sample & $N$ & $\beta$ & $\beta_{90}-\beta_0$ & $2\Delta\ln\mathcal{L}_\mathrm{cl}$ \\
\hline
All         & 61,938 & $-0.45\pm0.32$ & $+0.71\pm0.63$ & 10.9 \\
Pairs       & 43,478 & $-0.46\pm0.35$ & $+0.93\pm0.69$ & 7.1 \\
$>5''$      & 39,480 & $-0.72\pm0.33$ & $+0.26\pm0.65$ & 3.4 \\
$>10''$ (baseline) & 33,298 & $-0.48\pm0.32$ & $-0.08\pm0.63$ & 3.2 \\
\hline
Single comp. & 18,460 & $-0.50\pm0.82$ & $-0.66\pm1.78$ & -- \\
$\le10''$ or single & 28,640 & $-0.62\pm0.92$ & $+4.8\pm1.7$ & -- \\
\hline
\end{tabular}}
\tablefoot{\emph{All}: every component that lies above the \polsnr\ cut. \emph{Pairs}: islands in which both components pass the cut; $>5''$, $>10''$: pairs with larger component separation. The last is our baseline. $\beta_{90}-\beta_0$ is the peak difference with independent centres. $2\Delta\ln\mathcal{L}_\mathrm{cl}$ is the summed gain from freeing the centres in classes 7, 8 and 9. Uncertainties are computed from the likelihood curvature.}

\end{table}

The separation cut matters for the consistency of the two populations but not for $\beta$ (Table~\ref{tab:sep}). Without the cut, meaning for all 61,938 components that are above the \polsnr\ cut, $\beta=-0.45^\circ\pm0.32^\circ$ is the same. However, the peaks differ within classes by $2.0\sigma$, $0.8\sigma$ and $2.5\sigma$, with the sign in class 9 opposite to classes 7 and 8.  The origin is the closely spaced pairs. Pairs closer than $5''$, all of class 9, show no alignment at all ($\kappa\approx0.5$), implying that their $\psi$ carries no information on the source axis. Pairs at 5--$10''$ belong again almost all to class 9 and show no net alignment either, so their peak centres are poorly constrained. Requiring a separation of more than $10''$ removes the class-9 peak difference. 
Classes 7 and 8 are almost unaffected because nearly all their pairs are wider than $10''$. The joint $\beta$ changes by less than $0.25^\circ$ in all subsamples, and the excluded components give consistent values. 
Before the separation cut, the sample also contains 18,460 components whose partner in the island falls below the \polsnr\ cut.

These components can be included since $\psi$ is still measured between the two components of the island. Applying the same separation cut of $10''$ leaves $18\,167$ of them, which give $\beta=-0.62^\circ\pm0.82^\circ$. Combined with the baseline, $\beta=-0.50^\circ\pm0.30^\circ$. The gain in precision is negligible because these faint components lie mostly in the uniform background, so we keep pairs as the baseline.
These favour the perpendicular peak, which they populate more than twice as strongly as the standard-angle peak, whereas components in pairs populate both peaks equally.

The separation cut is motivated by the definition of $\psi$: it requires the two components to be clearly resolved from each other. Since the components are more than half a beam apart, every source also spans at least about 1.5 beams, so the cut takes the place of the angular-size cut of \citet{Trivedi2026}. We applied it after the analysis of the full sample. Of the four samples in Table~\ref{tab:sep}, $\beta$ differs by at most $0.27^\circ$, and the baseline value lies within this range, so this choice does not exaggerate the significance of the result.

\begin{figure}
\centering
\includegraphics[width=0.95\columnwidth]{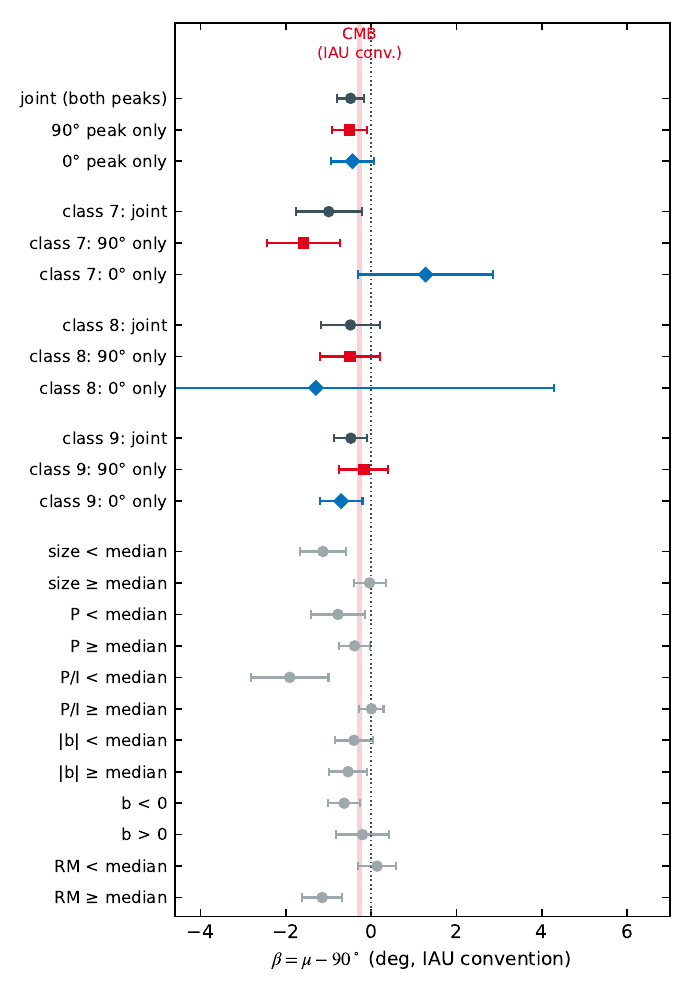}
\caption{$\beta$ (IAU convention) from the joint fit and from each peak alone, for the baseline sample and the individual classes, and for median splits in angular size, polarised intensity $P$, fractional polarisation $P/I$, $|b|$, Galactic hemisphere and signed RM. The splits are robustness tests carried out after the main analysis. The band shows the ACT+WMAP/\textit{Planck} value \citep{DiegoPalazuelos2025} converted to the IAU convention.}
\label{fig:forest}
\end{figure}

After the main analysis, we carried out a number of robustness tests.  Leaving out class 7 or class 8 changes $\beta$ by at most $0.08^\circ$. Leaving out class 9, which contains 70\% of the components, gives $\beta=-0.82^\circ\pm0.53^\circ$, consistent with the expected scatter ($0.8\sigma$). We also split the sample at the median of angular size, polarised intensity $P$, fractional polarisation $P/I$, $|b|$ and signed RM, and by Galactic hemisphere (Fig.~\ref{fig:forest}). All sources lie at $|b|>10^\circ$, so the hemisphere split compares 24,582 components at $b<-10^\circ$ with 8,716 at $b>10^\circ$. The $P$, $|b|$ and hemisphere splits agree within $0.6\sigma$. The size, $P/I$ and RM splits differ by $1.7\sigma$, $2.0\sigma$ and $2.0\sigma$. With six splits, differences of this size are expected: for independent splits, the largest of six exceeds $2\sigma$ in 26\% of cases, and the six differences together give $\chi^2=11.5$ for six degrees of freedom.  Because the likelihood depends on $\Delta-\mu$ only, re-running the same analysis with a hidden offset added to $\Delta$ returns the same result, and we verified this. 


\section{Systematic effects}
\label{sec:systematics}
First we will deal with an issue that is present in the pre-release DR1 catalogue, but will be fixed for the public version of Data Release 1. In the POSSUM DR1 catalogue, the polarisation angles $\chi$ and $\chi_0$ (and Stokes $Q$ and $U$) are measured from the $+y$ axis of each field image, that is, from north at the field centre $(\alpha_0,\delta_0)$ rather than from north at the position of the source. For an image in SIN projection, the angle to be added is

\begin{equation}
\begin{split}
\gamma = \arctan\bigl[&\sin\delta\,\sin(\alpha-\alpha_0),\\
  &\cos\delta\cos\delta_0+\sin\delta\sin\delta_0\cos(\alpha-\alpha_0)\bigr],
\end{split}
\label{eq:gamma}
\end{equation}
where $\arctan$ denotes the four-quadrant inverse tangent. For small offsets this is $\gamma\simeq(\alpha-\alpha_0)\sin\delta$. The structural angle $\psi$, computed from the component positions, is not affected. We reconstructed the centre of each of the 314 fields from the beam positions given in the catalogue and added $\gamma$ to every $\chi_0$. The correction improves the fit. Because $\gamma$ changes sign across each field, it mainly broadens the peaks, and it changes $\beta$ by only $+0.06^\circ$. Components flagged as lying in overlap regions between fields behave in the same way, so their angles also appear to come from single fields. All results in this paper use the corrected angles.

Each field has its own angle calibration, so the sample could contain one miscalibration angle per field. Random offsets between fields broaden the peaks and are absorbed into $\kappa$. We find that only their mean is degenerate with $\beta$, which is the analogue of the miscalibration angle in CMB analyses \citep{MinamiKomatsu2020}. POSSUM angles are accurate to about $\pm5^\circ$ per field \citep{Thomson2026}. We measured the scatter between fields directly by fitting $\mu$ separately in each of the 312 fields, with the other parameters fixed. The scatter beyond the statistical errors is $\tau=2.0^\circ$, consistent with zero, with $\tau<3.3^\circ$ at 95\% confidence. Random offsets of this size contribute at most $\tau/\sqrt{N_\mathrm{f}}\approx0.2^\circ$ to $\beta$. An offset common to all fields cannot be detected this way and remains fully degenerate with $\beta$. Random RM errors broaden the peaks and are absorbed by the error convolution, but an RM-correlated error, for example an inconsistent reference wavelength, would produce the RM dependence hinted at above. We find the de-rotation to be internally consistent to $0.004^\circ$.
We therefore treat the common offset as a nuisance angle $\alpha$, with $\beta_\mathrm{obs}=\beta+\alpha$ and a Gaussian prior of width $\sigma_\mathrm{cal}=0.3^\circ$ \citep{Thomson2026} which exceeds the $\lesssim0.2^\circ$ allowed by the random offsets between fields.

Finally, for barely resolved doubles, $\psi$ may not faithfully indicate the source axis. Pairs separated by less than a quarter of the beam indeed show no alignment, and pairs at a quarter to half a beam give discrepant peak centres (Sect.~\ref{sec:results}). Our baseline cut at half the beam removes both effects without changing $\beta$. 
 
 CMB analyses define $\beta>0$ as a clockwise rotation as seen by the observer \citep{Komatsu2022}, opposite to the IAU sense, so $\beta_\mathrm{CMB}=-\beta_\mathrm{IAU}$.  A positive $\beta_\mathrm{IAU}$ is therefore a counter-clockwise rotation on the sky. Because the comparison below hinges on this sign, we verified the full chain. In the catalogue, $\chi=\tfrac12\arctan(U,Q)$ to within $10^{-5}$\,deg, and ASKAP Stokes $Q$ and $U$ follow the IAU convention.
 The de-rotated angle equals $\chi-\mathrm{RM}\,\lambda_\mathrm{ref}^2$ to within $0.004^\circ$, with $\lambda_\mathrm{ref}=c/\nu_\mathrm{ref}$ the wavelength at the per-source reference frequency.
 The structural angle $\psi$ is measured from north through east for all 40,199 islands of the parent sample. We then rotated $Q+\mathrm{i}U$ by known angles $\beta_\mathrm{inj}=\pm0.5^\circ$, $\pm2^\circ$ and $+5^\circ$ and re-ran the full pipeline: RM de-rotation, $\Delta$ and the likelihood fit. The recovered shifts equal the injected rotations to within $0.001^\circ$, with the same sign.

\section{Joint analysis with the CMB}
\label{sec:cmb}

In the CMB convention our profile likelihood gives $\beta_\mathrm{RG}=+0.48^\circ \pm0.35^\circ$, $1.4\sigma$ from zero. Because the radio and CMB measurements are independent, we multiply our profile likelihood with Gaussian likelihoods for the published CMB results (Table~\ref{tab:cmb}). The CMB results already include their calibration priors. For the radio galaxies, we show results with the statistical error only and with the calibration offset marginalised. Because our profile likelihood is close to Gaussian (Fig.~\ref{fig:hist}) and $\alpha$ is independent of the CMB calibration, marginalising over $\alpha$ adds the two variances, $\sigma_\mathrm{RG}=(0.35^2+0.3^2)^{1/2}=0.46^\circ$.

The radio result agrees with every CMB value to within $0.6$--$0.8\sigma$. Combined with the joint ACT and \textit{Planck} analysis \citep{Eskilt2026}, $\beta=0.28^\circ\pm0.06^\circ$, $5.0\sigma$ from zero. The radio galaxies contribute only about 4\% of the weight with respect to ACT DR6. Their value lies in confirming the sign and amplitude with an independent method rather than in adding significance. The combination assumes that both probes measure the same angle. The CMB records the rotation since recombination, our sources lie at $z\lesssim2$. 

The assumption holds if an axion-like field rolls at late times, as in dark-energy models \citep{Choi2021,2026PhRvL.136d1004N,Minotti2026}. It also holds for ultralight axion dark matter that oscillates rapidly at last scattering. There the field averages to zero, which washes out the rotation imprinted at recombination \citep{Fedderke2019}. Both the CMB angle and the mean angle of a large radio-galaxy sample are then set by the field value today, and source-to-source oscillation phases only broaden the peaks. The assumption fails if the field evolved only before the emission of our sources, in which case radio galaxies should find $\beta\approx0$ \citep{Nakatsuka2022,2026arXiv261002667Y}.

\begin{table}
\caption{Radio galaxies (RG) combined with CMB results (CMB convention in degrees).}
\label{tab:cmb}
\centering
\small
\resizebox{\columnwidth}{!}{%
\begin{tabular}{lccc}
\hline\hline
CMB result & $\beta_\mathrm{CMB}$ & RG+CMB (stat.) & RG+CMB (+syst.) \\
\hline
ACT DR6$^a$          & $0.215\pm0.074$ & $0.226\pm0.072$ & $0.222\pm0.073$ \\
ACT+WMAP/\textit{Planck}$^b$ & $0.264\pm0.058$ & $0.270\pm0.057$ & $0.267\pm0.058$ \\
ACT+Cosmoglobe$^b$   & $0.231\pm0.059$ & $0.238\pm0.058$ & $0.235\pm0.059$ \\
ACT$\times$\textit{Planck}$^c$ & $0.277\pm0.057$ & $0.282\pm0.056$ & $0.280\pm0.057$ \\
\hline
\end{tabular}}
\tablefoot{$^a$\citet{DiegoPalazuelos2025}, $^b$combination given in \citet{DiegoPalazuelos2025} with \citet{EskiltKomatsu2022} or \citet{Eskilt2023}, $^c$\citet{Eskilt2026}. Significances of the combinations are: 3.1, 4.7, 4.1 and $5.0\sigma$.}
\end{table}
\section{Conclusions}

We have tested a sample of extended radio galaxies from the POSSUM sample for signs of birefringence. We have found the following:

(i) In POSSUM double sources, a population polarised parallel to the source axis is as abundant as the classical standard crosses and has the same intrinsic dispersion of $\sim 16^\circ$. (ii) Tying both to a common rotation in an unbinned von-Mises likelihood gives $\beta=-0.48^\circ\pm0.35^\circ$ (IAU) from 33,298 components of sources wider than half the beam, 2.0 times more precise than before. (iii) Components closer than half a beam carry no information on the source axis and shift the two peaks in a class-dependent way. Excluding them leaves $\beta$ unchanged and makes the two populations consistent in every class. (iv) In the CMB convention our result agrees with ACT and \textit{Planck} in sign and amplitude. 

Considering cosmic birefringence to arise from a late-time rotation, the combined value is $\beta=0.28^\circ\pm0.06^\circ$. At $0.35^\circ$ statistical precision, radio tests are limited by the absolute angle calibration. With a per-field calibration analysis, the completed POSSUM survey can test the CMB result independently at low redshift.\\

Extraordinary claims require extraordinary evidence. One way to amass more evidence is to collect redshifts for the host galaxies that would extend this test considerably. A per-field angle offset does not depend on redshift, so the difference in $\beta$ between redshift bins is insensitive to the absolute calibration. The dependence $\beta(z)$ also distinguishes between physical models. A constant Lorentz-violating $p_\mu$ produces a rotation that grows with distance and would constrain $p_0$ \citep{Carroll1990, Trivedi2026}. For an axion-like field, $\beta(z)$ traces the evolution of the field since $z\sim2$. Comparing it to the CMB then shows which fraction of the total rotation occurred at late times. A field that rolled near recombination predicts $\beta\approx0$ at all radio redshifts. In addition, host identifications would remove chance pairs, which populate the uniform background and would allow the size dependence to be tested with physical rather than angular sizes.

\begin{acknowledgements}
We thank E. Komatsu for helpful comments on the paper and B. Gaenssler for pointing us to the necessity of correcting the polarisation angles. MB acknowledges support by the DFG under Germany's Excellence Strategy -- EXC 2121 `Quantum Universe' -- 390833306 and by the DFG Research Group `Relativistic Jets' FOR5195 -- 443220636. SPO acknowledges support from the Comunidad de Madrid Atracci\'on de Talento program via grant 2022-T1/TIC-23797, and grant PID2023-146372OB-I00 funded by MICIU/AEI/10.13039/501100011033 and by ERDF, EU.
FN thanks the Forefront Physics and Mathematics Program to Drive Transformation (FoPM), a World-leading Innovative Graduate Study (WINGS) Program, the University of Tokyo. FN is also supported by JSPS KAKENHI Grant Number 24KJ0668, 20H05850, 23K20035, The ANRI Fellowship, and PID2023-146372OB-I00, funded by MICIU/AEI/10.13039/501100011033.
MR acknowledges support by the European Union – Next Generation EU and by the Italian Ministry of University and Research (MUR) via the PRIN 2022 project n. 20228WHTYC - CUP D53C24003550006 and from the research grant TAsP {\sl (Theoretical Astroparticle Physics)} funded by Istituto Nazionale di Fisica Nucleare (INFN). 
This scientific work uses data obtained from Inyarrimanha Ilgari Bundara / the Murchison Radio-astronomy Observatory. We acknowledge the Wajarri Yamaji People as the Traditional Owners and native title holders of the Observatory site. The Australian SKA Pathfinder is part of the Australia Telescope National Facility  which is managed by CSIRO. Operation of ASKAP is funded by the Australian Government with support from the National Collaborative Research Infrastructure Strategy. ASKAP uses the resources of the Pawsey Supercomputing Centre. Establishment of ASKAP, the Murchison Radio-astronomy Observatory and the Pawsey Supercomputing Centre are initiatives of the Australian Government, with support from the Government of Western Australia and the Science and Industry Endowment Fund. The POSSUM project has been made possible through funding from the Australian Research Council, the Natural Sciences and Engineering Research Council of Canada, the Canada Research Chairs Program, and the Canada Foundation for Innovation.

\end{acknowledgements}

\bibliographystyle{aa}
\bibliography{biref_twopop}

@article{Carroll1990,
  author  = {{Carroll}, S.~M. and {Field}, G.~B. and {Jackiw}, R.},
  title   = {Limits on a {Lorentz}- and parity-violating modification of electrodynamics},
  journal = {Phys. Rev. D},
  volume  = {41},
  pages   = {1231},
  year    = {1990}
}

@article{CarrollField1991,
  author  = {{Carroll}, S.~M. and {Field}, G.~B.},
  title   = {{Einstein} equivalence principle and the polarization of radio galaxies},
  journal = {Phys. Rev. D},
  volume  = {43},
  pages   = {3789},
  year    = {1991}
}

@article{Carroll1998,
  author  = {{Carroll}, S.~M.},
  title   = {Quintessence and the rest of the world: suppressing long-range interactions},
  journal = {Phys. Rev. Lett.},
  volume  = {81},
  pages   = {3067},
  year    = {1998}
}

@article{Harari1992,
  author  = {{Harari}, D. and {Sikivie}, P.},
  title   = {Effects of a {Nambu-Goldstone} boson on the polarization of radio galaxies and the cosmic microwave background},
  journal = {Phys. Lett. B},
  volume  = {289},
  pages   = {67},
  year    = {1992}
}

@article{KosteleckyMewes2008,
  author  = {{Kosteleck{\'y}}, V.~A. and {Mewes}, M.},
  title   = {Astrophysical tests of {Lorentz} and {CPT} violation with photons},
  journal = {ApJ},
  volume  = {689},
  pages   = {L1},
  year    = {2008}
}

@article{Choi2021,
  author  = {{Choi}, G. and {Lin}, W. and {Visinelli}, L. and {Yanagida}, T.~T.},
  title   = {Cosmic birefringence and electroweak axion dark energy},
  journal = {Phys. Rev. D},
  volume  = {104},
  pages   = {L101302},
  year    = {2021}
}

@ARTICLE{Fujita2021,
       author = {{Fujita}, Tomohiro and {Minami}, Yuto and {Murai}, Kai and {Nakatsuka}, Hiromasa},
        title = "{Probing axionlike particles via cosmic microwave background polarization}",
      journal = {\prd},
         year = 2021,
        month = mar,
       volume = {103},
       number = {6},
          eid = {063508},
        pages = {063508},
          doi = {10.1103/PhysRevD.103.063508},
       adsurl = {https://ui.adsabs.harvard.edu/abs/2021PhRvD.103f3508F}
}

@article{MinamiKomatsu2020,
  author  = {{Minami}, Y. and {Komatsu}, E.},
  title   = {New extraction of the cosmic birefringence from the {Planck} 2018 polarization data},
  journal = {Phys. Rev. Lett.},
  volume  = {125},
  pages   = {221301},
  year    = {2020}
}

@article{EskiltKomatsu2022,
  author  = {{Eskilt}, J.~R. and {Komatsu}, E.},
  title   = {Improved constraints on cosmic birefringence from the {WMAP} and {Planck} cosmic microwave background polarization data},
  journal = {Phys. Rev. D},
  volume  = {106},
  pages   = {063503},
  year    = {2022}
}

@ARTICLE{Louis2025,
       author = {{Louis}, Thibaut and {La Posta}, Adrien and {Atkins}, Zachary and {Jense}, Hidde T. and {Abril-Cabezas}, Irene and {Addison}, Graeme E. and {Ade}, Peter A.~R. and {Aiola}, Simone and {Alford}, Tommy and {Alonso}, David and {Amiri}, Mandana and {An}, Rui and {Austermann}, Jason E. and {Barbavara}, Eleonora and {Battaglia}, Nicholas and {Battistelli}, Elia Stefano and {Beall}, James A. and {Bean}, Rachel and {Beheshti}, Ali and {Beringue}, Benjamin and {Bhandarkar}, Tanay and {Biermann}, Emily and {Bolliet}, Boris and {Bond}, J. Richard and {Calabrese}, Erminia and {Capalbo}, Valentina and {Carrero}, Felipe and {Chen}, Shi-Fan and {Chesmore}, Grace and {Cho}, Hsiao-mei and {Choi}, Steve K. and {Clark}, Susan E. and {Cothard}, Nicholas F. and {Coughlin}, Kevin and {Coulton}, William and {Crichton}, Devin and {Crowley}, Kevin T. and {Darwish}, Omar and {Devlin}, Mark J. and {Dicker}, Simon and {Duell}, Cody J. and {Duff}, Shannon M. and {Duivenvoorden}, Adriaan J. and {Dunkley}, Jo and {Dunner}, Rolando and {Embil Villagra}, Carmen and {Fankhanel}, Max and {Farren}, Gerrit S. and {Ferraro}, Simone and {Foster}, Allen and {Freundt}, Rodrigo and {Fuzia}, Brittany and {Gallardo}, Patricio A. and {Garrido}, Xavier and {Gerbino}, Martina and {Giardiello}, Serena and {Gill}, Ajay and {Givans}, Jahmour and {Gluscevic}, Vera and {Goldstein}, Samuel and {Golec}, Joseph E. and {Gong}, Yulin and {Guan}, Yilun and {Halpern}, Mark and {Harrison}, Ian and {Hasselfield}, Matthew and {Healy}, Erin and {Henderson}, Shawn and {Hensley}, Brandon and {Herv{\'\i}as-Caimapo}, Carlos and {Hill}, J. Colin and {Hilton}, Gene C. and {Hilton}, Matt and {Hincks}, Adam D. and {Hlo{\v{z}}ek}, Ren{\'e}e and {Ho}, Shuay-Pwu Patty and {Hood}, John and {Hornecker}, Erika and {Huber}, Zachary B. and {Hubmayr}, Johannes and {Huffenberger}, Kevin M. and {Hughes}, John P. and {Ikape}, Margaret and {Irwin}, Kent and {Isopi}, Giovanni and {Joshi}, Neha and {Keller}, Ben and {Kim}, Joshua and {Knowles}, Kenda and {Koopman}, Brian J. and {Kosowsky}, Arthur and {Kramer}, Darby and {Kusiak}, Aleksandra and {Lagu{\"e}}, Alex and {Lakey}, Victoria and {Lee}, Eunseong and {Li}, Yaqiong and {Li}, Zack and {Limon}, Michele and {Lokken}, Martine and {Lungu}, Marius and {MacCrann}, Niall and {MacInnis}, Amanda and {Madhavacheril}, Mathew S. and {Maldonado}, Diego and {Maldonado}, Felipe and {Mallaby-Kay}, Maya and {Marques}, Gabriela A. and {van Marrewijk}, Joshiwa and {McCarthy}, Fiona and {McMahon}, Jeff and {Mehta}, Yogesh and {Menanteau}, Felipe and {Moodley}, Kavilan and {Morris}, Thomas W. and {Mroczkowski}, Tony and {Naess}, Sigurd and {Namikawa}, Toshiya and {Nati}, Federico and {Nerval}, Simran K. and {Newburgh}, Laura and {Nicola}, Andrina and {Niemack}, Michael D. and {Nolta}, Michael R. and {Orlowski-Scherer}, John and {Pagano}, Luca and {Page}, Lyman A. and {Pandey}, Shivam and {Partridge}, Bruce and {Perez Sarmiento}, Karen and {Prince}, Heather and {Puddu}, Roberto and {Qu}, Frank J. and {Ragavan}, Damien C. and {Ried Guachalla}, Bernardita and {Rogers}, Keir K. and {Rojas}, Felipe and {Sakuma}, Tai and {Schaan}, Emmanuel and {Schmitt}, Benjamin L. and {Sehgal}, Neelima and {Shaikh}, Shabbir and {Sherwin}, Blake D. and {Sierra}, Carlos and {Sievers}, Jon and {Sif{\'o}n}, Crist{\'o}bal and {Simon}, Sara and {Sonka}, Rita and {Spergel}, David N. and {Staggs}, Suzanne T. and {Storer}, Emilie and {Surrao}, Kristen and {Switzer}, Eric R. and {Tampier}, Niklas and {Thornton}, Robert and {Trac}, Hy and {Tucker}, Carole and {Ullom}, Joel and {Vale}, Leila R. and {Van Engelen}, Alexander and {Van Lanen}, Jeff and {Vargas}, Cristian and {Vavagiakis}, Eve M. and {Wagoner}, Kasey and {Wang}, Yuhan and {Wenzl}, Lukas and {Wollack}, Edward J. and {Zheng}, Kaiwen and {The Atacama Cosmology Telescope collaboration}},
        title = "{The Atacama Cosmology Telescope: DR6 power spectra, likelihoods and {\ensuremath{\Lambda}}CDM parameters}",
      journal = {\jcap},
         year = 2025,
        month = nov,
       volume = {2025},
       number = {11},
          eid = {062},
        pages = {062},
          doi = {10.1088/1475-7516/2025/11/062},
archivePrefix = {arXiv},
       eprint = {2503.14452},
 primaryClass = {astro-ph.CO},
       adsurl = {https://ui.adsabs.harvard.edu/abs/2025JCAP...11..062L}
}

@article{Komatsu2022,
  author  = {{Komatsu}, E.},
  title   = {New physics from the polarized light of the cosmic microwave background},
  journal = {Nat. Rev. Phys.},
  volume  = {4},
  pages   = {452},
  year    = {2022}
}

@article{HamakerBregman1996,
  author  = {{Hamaker}, J.~P. and {Bregman}, J.~D.},
  title   = {Understanding radio polarimetry. {III}. {Interpreting} the {IAU/IEEE} definitions of the {Stokes} parameters},
  journal = {A\&AS},
  volume  = {117},
  pages   = {161},
  year    = {1996}
}

@article{HavesConway1975,
  author  = {{Haves}, P. and {Conway}, R.~G.},
  title   = {The orientation of the magnetic field in radio sources},
  journal = {MNRAS},
  volume  = {173},
  pages   = {53P},
  year    = {1975}
}

@article{Clarke1980,
  author  = {{Clarke}, J.~N. and {Kronberg}, P.~P. and {Simard-Normandin}, M.},
  title   = {The relative orientation of magnetic fields and the radio axes in extended extragalactic radio sources},
  journal = {MNRAS},
  volume  = {190},
  pages   = {205},
  year    = {1980}
}

@article{SaikiaSalter1988,
  author  = {{Saikia}, D.~J. and {Salter}, C.~J.},
  title   = {Polarization properties of extragalactic radio sources},
  journal = {ARA\&A},
  volume  = {26},
  pages   = {93},
  year    = {1988}
}

@article{ListerHoman2005,
  author  = {{Lister}, M.~L. and {Homan}, D.~C.},
  title   = {{MOJAVE}: Monitoring of Jets in Active Galactic Nuclei with {VLBA} Experiments. {I}. {First}-epoch 15 {GHz} linear polarization images},
  journal = {AJ},
  volume  = {130},
  pages   = {1389},
  year    = {2005}
}

@article{Battye2026,
  author  = {{Battye}, R.~A. and {Jackson}, N. and {Browne}, I.~W.~A.},
  title   = {Limits on global cosmic birefringence using radio sources},
  journal = {arXiv e-prints},
  eprint  = {2607.02664},
  year    = {2026}
}

@article{Trivedi2026,
  author  = {{Trivedi}, P. and {Dreismann}, J. and {Br{\"u}ggen}, M. and others},
  title   = {Cosmic birefringence test of relativistic electrodynamics with {POSSUM} radio galaxies},
  journal = {in prep.},
  year    = {2026}
}

@article{Hotan2021,
  author  = {{Hotan}, A.~W. and {Bunton}, J.~D. and {Chippendale}, A.~P. and others},
  title   = {{Australian Square Kilometre Array Pathfinder}: {I}. {System} description},
  journal = {PASA},
  volume  = {38},
  pages   = {e009},
  year    = {2021}
}

@article{Gaensler2025,
  author  = {{Gaensler}, B.~M. and others},
  title   = {The {Polarisation Sky Survey of the Universe's Magnetism} ({POSSUM}): science goals and survey description},
  journal = {PASA},
  volume  = {42},
  pages   = {e91},
  year    = {2025}
}

@article{Norris2021,
  author  = {{Norris}, R.~P. and others},
  title   = {The {Evolutionary Map of the Universe} pilot survey},
  journal = {PASA},
  volume  = {38},
  pages   = {e046},
  year    = {2021}
}

@article{Whiting2012,
  author  = {{Whiting}, M. and {Humphreys}, B.},
  title   = {Source-finding for the {Australian Square Kilometre Array Pathfinder}},
  journal = {PASA},
  volume  = {29},
  pages   = {371},
  year    = {2012}
}

@ARTICLE{Perez2025,
       author = {{Baron Perez}, Nicolas and {Br{\"u}ggen}, Marcus and {Kasieczka}, Gregor and {Lucie-Smith}, Luisa},
        title = "{Classification of radio sources through self-supervised learning}",
      journal = {\aap},
         year = 2025,
        month = jul,
       volume = {699},
          eid = {A302},
        pages = {A302},
          doi = {10.1051/0004-6361/202554735},
archivePrefix = {arXiv},
       eprint = {2503.19111},
 primaryClass = {astro-ph.IM},
       adsurl = {https://ui.adsabs.harvard.edu/abs/2025A&A...699A.302B}
}

@book{MardiaJupp2000,
  author    = {{Mardia}, K.~V. and {Jupp}, P.~E.},
  title     = {Directional Statistics},
  publisher = {Wiley},
  address   = {Chichester},
  year      = {2000}
}

@article{vonMises1918,
  author  = {{von Mises}, R.},
  title   = {{\"U}ber die ``{Ganzzahligkeit}'' der {Atomgewichte} und verwandte {Fragen}},
  journal = {Physikalische Zeitschrift},
  volume  = {19},
  pages   = {490},
  year    = {1918}
}

@article{BestFisher1981,
  author  = {{Best}, D.~J. and {Fisher}, N.~I.},
  title   = {The bias of the maximum likelihood estimators of the von {Mises-Fisher} concentration parameters},
  journal = {Commun. Stat. Simul. Comput.},
  volume  = {10},
  pages   = {493},
  year    = {1981}
}

@ARTICLE{DiegoPalazuelos2025,
       author = {{Diego-Palazuelos}, P. and {Komatsu}, E.},
        title = "{Cosmic Birefringence from the Atacama Cosmology Telescope Data Release 6}",
      journal = {arXiv e-prints},
         year = 2025,
        month = sep,
          eid = {arXiv:2509.13654},
        pages = {arXiv:2509.13654},
          doi = {10.48550/arXiv.2509.13654},
archivePrefix = {arXiv},
       eprint = {2509.13654},
 primaryClass = {astro-ph.CO},
       adsurl = {https://ui.adsabs.harvard.edu/abs/2025arXiv250913654D}
}

@article{Eskilt2026,
  author  = {{Eskilt}, J.~R.},
  title   = {Cosmic birefringence from a joint analysis of {ACT} and {Planck}},
  journal = {arXiv e-prints},
  eprint  = {2608.06480},
  year    = {2026}
}

@ARTICLE{Eskilt2023,
       author = {{Eskilt}, J.~R. and {Watts}, D.~J. and {Aurlien}, R. and {Basyrov}, A. and {Bersanelli}, M. and {Brilenkov}, M. and {Colombo}, L.~P.~L. and {Eriksen}, H.~K. and {Fornazier}, K.~S.~F. and {Franceschet}, C. and {Fuskeland}, U. and {Galloway}, M. and {Gjerl{\o}w}, E. and {Hensley}, B. and {Hergt}, L.~T. and {Herman}, D. and {Ihle}, H.~T. and {Lee}, K. and {Lunde}, J.~G.~S. and {Nerval}, S.~K. and {Paradiso}, S. and {Patel}, S.~K. and {Rahman}, F. and {Regnier}, M. and {San}, M. and {Sanyal}, S. and {Stutzer}, N.-O. and {Thommesen}, H. and {Verma}, A. and {Wehus}, I.~K. and {Zhou}, Y.},
        title = "{COSMOGLOBE DR1 results. II. Constraints on isotropic cosmic birefringence from reprocessed WMAP and Planck LFI data}",
      journal = {\aap},
         year = 2023,
        month = nov,
       volume = {679},
          eid = {A144},
        pages = {A144},
          doi = {10.1051/0004-6361/202346829},
archivePrefix = {arXiv},
       eprint = {2305.02268},
 primaryClass = {astro-ph.CO},
       adsurl = {https://ui.adsabs.harvard.edu/abs/2023A&A...679A.144E}
}

@ARTICLE{Thomson2026,
       author = {{Thomson}, Alec J.~M. and {Galvin}, Timothy J. and {Duchesne}, Stefan W. and {Lenc}, Emil and {Heald}, George and {Hlinka}, Ondrej and {Malik}, Sunil and {Anderson}, Craig S. and {Osinga}, Erik and {Baidoo}, Lerato and {McClure-Griffiths}, N.~M. and {Hutschenreuter}, Sebastian and {O'Sullivan}, Shane P. and {Akahori}, Takuya and {Gaensler}, B.~M. and {Leahy}, J.~P. and {Ma}, Y.~K. and {Moss}, Vanessa A. and {Rudnick}, L. and {Van Eck}, C.~L. and {West}, J.~L.},
        title = "{The Rapid ASKAP Continuum Survey VII: Spectra and Polarisation in Cutouts of Extragalactic sources (SPICE-RACS) second data release ─ unveiling the magnetised sky}",
      journal = {\pasa},
         year = 2026,
        month = jun,
       volume = {43},
          eid = {e089},
        pages = {e089},
          doi = {10.1017/pasa.2026.10214},
archivePrefix = {arXiv},
       eprint = {2605.16917},
 primaryClass = {astro-ph.GA},
       adsurl = {https://ui.adsabs.harvard.edu/abs/2026PASA...43...89T}
}

@ARTICLE{2026arXiv261002667Y,
       author = {{Yin}, Lu and {Komatsu}, Eiichiro},
        title = "{Distinguishing the origin of cosmic birefringence: dark energy, dark matter, and neutrino asymmetry}",
      journal = {arXiv e-prints},
         year = 2026,
        month = oct,
          eid = {arXiv:2610.02667},
        pages = {arXiv:2610.02667},
          doi = {10.48550/arXiv.2610.02667},
archivePrefix = {arXiv},
       eprint = {2610.02667},
 primaryClass = {astro-ph.CO},
       adsurl = {https://ui.adsabs.harvard.edu/abs/2026arXiv261002667Y}
}

@ARTICLE{2026PhRvL.136d1004N,
       author = {{Naokawa}, Fumihiro},
        title = "{Universal Profile for Cosmic Birefringence Tomography with Radio Galaxies}",
      journal = {\prl},
         year = 2026,
        month = jan,
       volume = {136},
       number = {4},
          eid = {041004},
        pages = {041004},
          doi = {10.1103/srfg-9fdy},
archivePrefix = {arXiv},
       eprint = {2504.06709},
 primaryClass = {astro-ph.CO},
       adsurl = {https://ui.adsabs.harvard.edu/abs/2026PhRvL.136d1004N}
}

@ARTICLE{Eskilt2022,
       author = {{Eskilt}, J.~R.},
        title = "{Frequency-dependent constraints on cosmic birefringence from the LFI and HFI Planck Data Release 4}",
      journal = {\aap},
         year = 2022,
        month = jun,
       volume = {662},
          eid = {A10},
        pages = {A10},
          doi = {10.1051/0004-6361/202243269},
archivePrefix = {arXiv},
       eprint = {2201.13347},
 primaryClass = {astro-ph.CO},
       adsurl = {https://ui.adsabs.harvard.edu/abs/2022A&A...662A..10E}
}

@ARTICLE{DiegoPalazuelos2022,
       author = {{Diego-Palazuelos}, P. and {Eskilt}, J.~R. and {Minami}, Y. and {Tristram}, M. and {Sullivan}, R.~M. and {Banday}, A.~J. and {Barreiro}, R.~B. and {Eriksen}, H.~K. and {G{\'o}rski}, K.~M. and {Keskitalo}, R. and {Komatsu}, E. and {Mart{\'\i}nez-Gonz{\'a}lez}, E. and {Scott}, D. and {Vielva}, P. and {Wehus}, I.~K.},
        title = "{Cosmic Birefringence from the Planck Data Release 4}",
      journal = {\prl},
         year = 2022,
        month = mar,
       volume = {128},
       number = {9},
          eid = {091302},
        pages = {091302},
          doi = {10.1103/PhysRevLett.128.091302},
archivePrefix = {arXiv},
       eprint = {2201.07682},
 primaryClass = {astro-ph.CO},
       adsurl = {https://ui.adsabs.harvard.edu/abs/2022PhRvL.128i1302D}
}

@ARTICLE{Nakatsuka2022,
       author = {{Nakatsuka}, Hiromasa and {Namikawa}, Toshiya and {Komatsu}, Eiichiro},
        title = "{Is cosmic birefringence due to dark energy or dark matter? A tomographic approach}",
      journal = {\prd},
         year = 2022,
        month = jun,
       volume = {105},
       number = {12},
          eid = {123509},
        pages = {123509},
          doi = {10.1103/PhysRevD.105.123509},
archivePrefix = {arXiv},
       eprint = {2203.08560},
 primaryClass = {astro-ph.CO},
       adsurl = {https://ui.adsabs.harvard.edu/abs/2022PhRvD.105l3509N}
}

@ARTICLE{Gardner1969,
       author = {{Gardner}, F.~F. and {Morris}, D. and {Whiteoak}, J.~B.},
        title = "{The linear polarization of radio sources between 11 and 20 cm wavelength. I. Observations.}",
      journal = {Australian Journal of Physics},
         year = 1969,
        month = jan,
       volume = {22},
        pages = {79-106},
          doi = {10.1071/PH690079},
       adsurl = {https://ui.adsabs.harvard.edu/abs/1969AuJPh..22...79G}
}

@ARTICLE{Namikawa2025,
       author = {{Namikawa}, Toshiya},
        title = "{Resolving the Negative Effective Neutrino Mass Parameter with Cosmic Birefringence}",
      journal = {\prl},
         year = 2025,
        month = oct,
       volume = {135},
       number = {16},
          eid = {161004},
        pages = {161004},
          doi = {10.1103/qgnn-6hsf},
archivePrefix = {arXiv},
       eprint = {2506.22999},
 primaryClass = {astro-ph.CO},
       adsurl = {https://ui.adsabs.harvard.edu/abs/2025PhRvL.135p1004N}
}

@ARTICLE{Minotti2026,
       author = {{Minotti}, Andrea and {Wu}, Yunzhi and {Zanzarella}, Tiziano and {Fornengo}, Nicolao and {Regis}, Marco and {Taoso}, Marco},
        title = "{Cosmic birefringence from a stiff axion field and its cosmological implications}",
      journal = {arXiv e-prints},
         year = 2026,
        month = oct,
          eid = {arXiv:2610.27082},
        pages = {arXiv:2610.27082},
          doi = {10.48550/arXiv.2610.27082},
archivePrefix = {arXiv},
       eprint = {2610.27082},
 primaryClass = {hep-ph},
       adsurl = {https://scixplorer.org/abs/2026arXiv261027082M}
}

@article{Fedderke2019,
  author  = {Fedderke, Michael A. and Graham, Peter W. and Rajendran, Surjeet},
  title   = {Axion dark matter detection with CMB polarization},
  journal = {Phys. Rev. D}, 
  volume = {100}, 
  pages = {015040}, 
  year = {2019},
  eprint  = {1903.02666}, doi = {10.1103/PhysRevD.100.015040}}

\end{document}